\documentclass[english,aps,preprint]{revtex4}
\usepackage{amsmath}
\usepackage{amssymb}

\makeatletter

\providecommand{\LyX}{L\kern-.1667em\lower.25em\hbox{Y}\kern-.125emX\@}

\makeatother

\usepackage{babel}

\begin{document}

\title{Chiral geometries of (2+1)-d AdS gravity}

\author{David A. Lowe and Shubho Roy}

\email{lowe@brown.edu, sroy@het.brown.edu}

\affiliation{Department of Physics, Brown University, Providence, RI 02912, USA}

\begin{abstract}
Pure gravity in (2+1)-dimensions with negative cosmological constant
is classically equivalent Chern-Simons gauge theory with gauge group
$SO(2,2)$, which may be realized on chiral and anti-chiral gauge
connections. This paper looks at half-AdS geometries i.e. those with
a trivial right-moving gauge connection while the left-moving connection
is a standard (Ba\~nados-Teitelboim-Zanelli) BTZ connection. These
are shown to be related by diffeomorphism to a BTZ geometry with different
mass and angular momentum. Generically this is over-spinning, leading
to a naked closed timelike curves. Other closely related solutions
are also studied. These results suggest that the measure of the Chern-Simons
path integral cannot factorize in a chiral way, if it is to represent
a sum over physically sensible states.
\end{abstract}
\maketitle

\section{introduction}

General Relativity in (2+1)-dimensions has long been considered an
arena for investigating issues in gravity without the additional complications
of gravitational dynamics in higher dimensions. The reason for this
is that in (2+1)-dimensions there are no local propagating degrees
of freedom in the bulk i.e. there are no gravitational waves in the
bulk and the dynamics arises purely from global topology - defects
and boundaries \cite{Deser:1983dr,Deser:1983tn} (for a comprehensive
review consult\cite{S.Carlip1998}). Although the theory does not
have gravitational waves i.e. no gravitational attraction, interestingly,
it does have black hole solutions \cite{Banados:1992gq,Banados:1992wn}
for the case of negative cosmological constant. These (BTZ) black
hole metrics are locally isomorphic to the maximally symmetric solution,
namely anti de Sitter space and share the same asymptotic symmetry
group - the conformal group in 2-dimensions \cite{Brown:1986nw}.
The infinite set of Virasoro charges of this asymptotic conformal
group parametrize all asymptotically AdS metrics\cite{Banados:1998gg}.
In particular for BTZ metrics, the Virasoro charges ($L_{0},\bar{L}_{0}$)
are simply linear combinations of the mass ($M$) and spin ($J$)
of the black hole ($L_{0}=\frac{Ml+J}{2},$ $\bar{L}_{0}=\frac{Ml-J}{2}$).

There is a long history of efforts formulating gravity in general
$d$-dimensions as a gauge theory by combining the vierbein and spin
connection into a single $ISO(d-1,1)$ gauge connection, since small
diffeomorphisms can be expressed as local Lorentz rotations and translations
\cite{MacDowell:1977jt} on shell. However this program was abortive
since the Einstein-Hilbert action could not be expressed in terms
of the gauge connection. This obstacle was circumvented for the $d=2+1$
case \cite{Achucarro:1989gm,Witten:1988hc} for general nonvanishing
cosmological constant and the Einstein-Hilbert action was expressed
as a Chern-Simons action for the $ISO(2,1)$ connection for zero cosmological
constant, and a $SO(2,2)$ connection for negative cosmological constant.
Since $SO(2,2)\equiv SL(2,\mathbb{R})\times SL(2,\mathbb{R})/\mathbb{Z}_{2}$
(the $\mathbb{Z}_{2}$ acts as $-1$ on the $SL(2,\mathbb{R})$ connections),
the action can be written as a sum of two Chern-Simons terms with
independent $SL(2,\mathbb{R})$ gauge connections. By adding an additional
topological term to the Einstein action\cite{Witten:1988hc} the
coefficients of the left/right Chern-Simons terms become independent.
Based on this motivation, it has been conjectured the path integral
for pure $2+1$-d gravity with negative cosmological constant factors
holomorphically \cite{Witten:2007kt,Maloney:2007ud,Gaberdiel:2007ve,Gaiotto:2007xh,Manschot:2007zb,Yin:2007at,Yin:2007gv,Avramis:2007gx}.
Also in other extensions of $2+1$-d AdS gravity with a gravitational
Chern-Simons terms there are sectors where the left (or right) gauge
connection can be pure gauge i.e. geometries that are chiral \cite{Li:2008dq,Carlip:2008jk}.
In such cases again the path integral is expected to be holomorphic
factorizable. Of course there are several issues - the classical equivalence
of Einstein's theory and the Chern-Simons theory might not extend
to the quantum realm in such a simple manner. This is certainly true
for large diffeomorphisms when the vierbein is noninvertible and the
metric interpretation is unclear. \\
 \\
 The aim of this paper is to study geometries where one of the CS
gauge fields is set to be globally trivial (anti-de Sitter) with the
other gauge field being nontrivial (BTZ-like). If holomorphic factorization
holds at the level of the Chern-Simons path integral, such geometries
should have a meaningful metric interpretation. We find that such
metrics generically have naked closed timelike curves (CTCs) and hence
do not respect causality. The plan of the paper is as follows. In
section 2 we review the Chern-Simons formulation of the BTZ solution
and note the relationship between the holonomies and the casimirs
(mass and spin). In section 3, we construct \char`\"{}hybrid'' metrics
made of a left($^{+}$) connection of a $(M,J)$ BTZ solution and
a right($^{-}$) connection of a $(m,\, j)$ BTZ solution. The hybrid
metric after a suitable change of coordinates is shown to be another
BTZ solution with charges $(\frac{M+m}{2}+\frac{J-j}{2l},\frac{(M-m)l}{2}+\frac{J+j}{2})$.
In section 4 we set the right connection to pure AdS$_{3}$ and note
that these geometries are super-rotating BTZ solutions which necessarily
have naked CTCs thus violating causality. The right connection when
set to zero instead of pure AdS$_{3}$ gives rise to singular metrics.

\section{Chern-Simons formulation of the BTZ black hole}

To start with, let us very briefly review the Chern Simons formulation
of Lorentzian $2+1$d gravity with negative cosmological constant
$\Lambda=-\frac{1}{l^{2}}$. The details can be found in \cite{Witten:1988hc,Banados:1998gg,Carlip:2008jk}.
The vierbein, $e^{a}$ and the spin connection one form $\omega^{a}$
are combined into a $SO(2,2)$ gauge connection one form, \[
A=e^{a}P_{a}+\omega^{a}J_{a}\]
with the algebra, \[
[J_{a},J_{b}]=\epsilon_{ab}^{c}J_{c}\,,\:[P_{a},P_{b}]=\frac{1}{l^{2}}\epsilon_{ab}^{c}J_{c}\,,\:[J_{a},P_{b}]=\epsilon_{ab}^{c}P_{c}\,.\]

The $SO(2,2)$ generators can be split into two $SL(2,\mathbb{R})$
copies, \[
T_{a}^{\pm}=\frac{1}{2}(J_{a}\pm lP_{a})\]
 satisfying\[
[T_{a}^{+},T_{b}^{+}]=\epsilon_{ab}\,^{c}T_{c}^{+}\,,\:[T_{a}^{-},T_{b}^{-}]=\epsilon_{ab}\,^{c}T_{c}^{-}\,,\:[T_{a}^{+},T_{b}^{-}]=0.\]
\[
\]
Here $\epsilon^{012}=1$ and $\eta=$ diag $(-1,1,1)$. The gauge
connection accordingly factorizes to, \begin{eqnarray*}
A & = & A^{a+}T_{a}^{+}+A^{a-}T_{a}^{-}\\
{}A^{\pm} & = & \omega\pm\frac{e}{l}.\end{eqnarray*}

The Einstein-Hilbert action with a negative constant term can then
be cast as the difference of two Chern-Simons terms, \[
I=I_{CS}[A^{+}]-I_{CS}[A^{-}],\]
 where the Chern-Simons action is, \[
I_{CS}[A]=\frac{k}{4\pi}\int{Tr(dA\wedge A+\frac{2}{3}A\wedge A\wedge A)},\]
 where the trace is over the 2-dimensional representation of $SL(2,\mathbb{R})$,
\[
T_{0}=-i\frac{\sigma^{2}}{2},\, T_{1}=-i\frac{\sigma^{3}}{2},\, T_{2}=\frac{\sigma^{1}}{2},\]
 and the Chern-Simons coupling constant or level number, \[
k=-2l.\]
 Following \cite{Banados:1992gq,Banados:1992wn,Banados:1998gg} we
choose the convention $8G=1$. The solutions to Einstein's equations
are given by flat connections \[
dA^{\pm}+A^{\pm}\wedge A^{\pm}=0.\]
 In fact the above pair of flat connection equations of motion can
also be obtained by any action which is the difference of two Chern-Simons
actions with \emph{different} Chern-Simons couplings i.e. $k^{+}\neq k^{-}$.
This corresponds to adding a topological term to the Einstein action
with a coupling proportional to the difference ($k^{+}-k^{-}$), which
does not change the local equations of motion, but will be important
if the full path integral is considered.

Flat connections are classified according to their holonomy, i.e.
every flat connection can be gauge transformed to the form \begin{equation}
A^{\pm}=U^{\pm-1}dU^{\pm}\label{eq:gtrivial}\end{equation}
where $U\in SL(2,\mathbb{R})$ is a gauge transformation element.
The connection is globally pure gauge only if $U$ is single valued.
The holonomy, $w$, for a connection is given by the Wilson loop operator
along noncontractible cycles \[
W^{\pm}=\exp(\oint A^{\pm})=\exp(w^{\pm})\,.\]

The local pure gauge condition is a reflection of the fact that in
2+1d any solution of Einstein equation with a negative cosmological
constant is locally anti-de Sitter space, which is the maximally symmetric
solution. Globally distinct solutions can be obtain by orbifolding
the maximally symmetric AdS$_{3}$ spacetime, i.e. identifying points
along an orbit of some killing vector, thus introducing nontrivial
cycles. This leads to black hole solutions \cite{Banados:1992gq,Banados:1992wn}
with mass $M$ and spin $J$ given by the metric, \[
ds^{2}=-N(r)^{2}dt^{2}+N(r)^{-}2dr^{2}+(r^{2}N^{\phi}dt+d\phi)^{2}\]
 where, \[
N^{2}(r)=-M+\frac{r^{2}}{l^{2}}+\frac{J^{2}}{4r^{2}}\]
 and \[
N^{\phi}=-\frac{J}{2r^{2}}\]
and $t\in(-\infty,\infty)$, $r\in(0,\infty)$ and $\phi\in[0,2\pi)$.
Now quotienting the maximally symmetric AdS$_{3}$ space reduces the
number of symmetries (or Killing vectors) by demanding bonafide tensor
fields respect single-valuedness after the periodic identifications.
The necessary and sufficient condition for which is that tensor fields
must commute with the Killing vectors belonging to the identification
subgroup (along the orbits of which the identifications were made).
This is true only for $\frac{\partial}{\partial t}$ and $\frac{\partial}{\partial\phi}$
for the BTZ. Thus the symmetry group of BTZ is not $SO(2,2)$ but
$R\times SO(1)$\cite{Banados:1992gq}.

The $SO(2,2)$ gauge connection was worked out in \cite{Cangemi:1992my}
but these have the unpleasant feature that the connections for the
extreme BTZ are singular. So we use a different gauge in which the
left and right connections appear as\[
\]
\[
A^{+0}=-N(r)\left(\frac{dt}{l}-d\phi\right)\,,\: A^{+1}=\frac{1-\frac{Jl}{2r^{2}}}{N(r)}\frac{dr}{l}\,,\: A^{+2}=-\left(\frac{J}{2r}+\frac{r}{l}\right)\left(\frac{dt}{l}-d\phi\right)\,,\]
\[
A^{-0}=-N(r)\left(\frac{dt}{l}+d\phi\right)\,,\: A^{-1}=-\frac{1+\frac{Jl}{2r^{2}}}{N(r)}\frac{dr}{l}\,,\: A^{-2}=\left(\frac{J}{2r}-\frac{r}{l}\right)\left(\frac{dt}{l}+d\phi\right)\]
 and the corresponding gauge transformation elements are obtained
by solving \eqref{eq:gtrivial} \begin{eqnarray*}
U^{\pm} & = & e^{\theta_{0\pm}T_{0}}e^{\theta_{1\pm}T_{1}}e^{\theta_{2\pm}T_{2}}\\
\theta_{0\pm} & = & 0\\
\sinh{\theta_{1\pm}} & = & \frac{N(r)}{\sqrt{{M\pm\frac{J}{l}}}}\\
\cosh{\theta_{1\pm}} & = & \frac{\frac{J}{2r}\pm\frac{r}{l}}{\sqrt{{M\pm\frac{J}{l}}}}\\
\theta_{2}{\pm} & = & \sqrt{M\pm\frac{J}{l}}\left(\phi\mp\frac{t}{l}\right).\end{eqnarray*}

Clearly these transformation are not suitable for the extreme case
$J=\pm Ml$. For $J=Ml$, the nonsingular gauge group element is,
\begin{eqnarray*}
U^{-}=e^{\theta(T_{0}+T_{2})}e^{\theta_{1}T_{1}},\\
\theta=-ln|\frac{r}{l}-\frac{Ml}{2r}|,\\
\theta_{1}=-\left(\phi+\frac{t}{l}\right).\end{eqnarray*}
 Finally the Wilson loop operator along a constant $t$, $\phi$ loop,
\[
W^{\pm}=\left(\begin{array}{cc}
\cosh{\pi\sqrt{M\pm\frac{J}{l}}} & e^{-}{\theta_{1}}\sinh{\pi\sqrt{M\pm\frac{J}{l}}}\\
e^{\theta_{1}}\sinh{\pi\sqrt{M\pm\frac{J}{l}}} & \cosh{\pi\sqrt{M\pm\frac{J}{l}}}\end{array}\right)\,.\]
 So the eigenvalues are $e^{\lambda}$, $\lambda=\pm\sqrt{M\pm\frac{J}{l}}$
and the holonomy matrices turn out to be, \[
w_{\pm}=\left(\begin{array}{cc}
\pm\pi\sqrt{M\pm\frac{J}{l}} & 0\\
0 & \mp\pi\sqrt{M\pm\frac{J}{l}}\end{array}\right)\]
 This gives the quadratic Casimirs, \begin{eqnarray*}
Tr(w_{+}^{2}+w_{-}^{2})=4\pi^{2}M\\
Tr(w_{+}^{2}-w_{-}^{2})=4\pi^{2}\frac{J}{l}\,.\end{eqnarray*}

\section{Hybrid Geometries}

In this section we are interested in metrics resulting from combining
the left and right $SL(2,\mathbb{R})$ connections for different BTZ
solutions, say we combine a left connection for a mass $M$, spin
$J$ and a right connection for mass $m$ and spin $j$ BTZ metric.
The metric for such a hybrid geometry is then, \[
ds^{2}=g_{tt}(r)dt^{2}+2g_{t\phi}(r)dtd\phi+g_{rr}(r)dr^{2}+g_{\phi\phi}(r)d\phi^{2},\]
 with, \begin{eqnarray}
\frac{g_{tt}}{l^{2}} & = & \frac{M+m}{4}+\frac{Jj}{8r^{2}}-\frac{r^{2}}{2l^{2}}-\frac{1}{2}N_{1}(r)N_{2}(r)\nonumber \\
\frac{g_{t\phi}}{l^{2}} & = & -\frac{j+J-ml+Ml}{4l}\nonumber \\
\frac{g_{rr}}{l^{2}} & = & \left(\frac{1-\frac{Jl}{2r^{2}}}{2N_{1}(r)}+\frac{1+\frac{jl}{2r^{2}}}{2N_{2}(r)}\right)^{2}\nonumber \\
\frac{g_{\phi\phi}}{l^{2}} & = & \frac{m+M}{4}-\frac{jJ}{8r^{2}}+\frac{J-j}{2l}+\frac{r^{2}}{2l^{2}}+\frac{1}{2}N_{1}(r)N_{2}(r)\label{eq:hybridmet}\end{eqnarray}
where $N_{1}(r)=\sqrt{-M+\frac{r^{2}}{l^{2}}+\frac{J^{2}}{4r^{2}}}$
and $N_{2}(r)=\sqrt{-m+\frac{r^{2}}{l^{2}}+\frac{j^{2}}{4r^{2}}}$.
In these coordinates the metric or the geometry looks highly unintuitive
since the metric components are quartic in $r$ and the determinant
seems to have multiple non-coincident zeroes making it hard to analyze
the location of horizons, surfaces bounding regions with CTCs, etc.
for the general $M$, $m$, $j$, $J$. To gain insight we switch
over to a different set of coordinates as follows. It has been shown
\cite{Banados:1998gg} that the most general solution to Einstein
equations with negative cosmological constant in (2+1)-dimensions
with asymptotically anti-de Sitter boundary conditions is of the form
\begin{equation}
ds^{2}=\frac{lL(w)}{2}dw^{2}+\frac{l\bar{L}(\bar{w})}{2}d{\bar{w}}^{2}+\left(l^{2}e^{2\rho}+\frac{L(w)\bar{L}(\bar{w})}{4}e^{-2\rho}\right)dwd\bar{w}+l^{2}d\rho^{2},\label{eq:stdmet}\end{equation}
for arbitrary functions $L(w)$ and $\bar{L}(\bar{w})$. Here $w=\phi+\frac{t}{l}$,
$\bar{w}=\phi-\frac{t}{l}$ are the boundary coordintes and $\rho$
is a radial coordinate. $\phi\in[0,2\pi)$ is an angular coordinate,
while $t$ is the time coordinate. In particular the BTZ solution
is a special case with constant metric coefficients, \begin{eqnarray*}
L(w)=L_{0}=\frac{Ml+J}{2}\\
\bar{L}(\bar{w})=\bar{L}_{0}=\frac{Ml-J}{2}\end{eqnarray*}

So after effecting the following change of coordinates, \begin{eqnarray*}
w=\phi+\frac{t}{l}\\
\bar{w}=\phi-\frac{t}{l}\end{eqnarray*}

\[
\rho=\frac{1}{2}\ln\left|\frac{\left(N_{1}+\frac{r}{l}+\frac{J}{2r}\right)\left(N_{2}+\frac{r}{l}-\frac{j}{2r}\right)}{4}\right|\]

we arrive at the metric in the form of \eqref{eq:stdmet} with, \begin{eqnarray}
L(w) & = & \frac{Ml+J}{2}\nonumber \\
\bar{L}(\bar{w}) & = & \frac{ml-j}{2}\,.\label{eq:hybridans}\end{eqnarray}

So it is just another BTZ black hole metric with ADM mass, $M_{ADM}$
and spin, $J_{ADM}$ given by, \begin{eqnarray*}
M_{ADM} & = & \frac{M+m}{2}+\frac{J-j}{2l}\\
J_{ADM} & = & \frac{(M-m)l}{2}+\frac{J+j}{2}\,.\end{eqnarray*}
 A crucial observation is that if the extremality bounds $Ml\geq|J|$
and $ml\geq|j|$ are satisfied then the resulting black hole also
satisfies an extremality bound, \begin{equation}
M_{ADM}^{2}l^{2}-J_{ADM}^{2}=(Ml+J)(ml-j)\geq0\,.\label{eq:extremality}\end{equation}

\section{Chiral geometries}

\subsection{Left BTZ - Right AdS$_{3}$ geometries}

For this case we set $m=-1$, $j=0$ i.e. pure AdS$_{3}$ space instead
of a black hole right connection. So the resulting black hole has
mass and spin, \begin{eqnarray*}
M_{ADM} & = & \frac{M-1}{2}+\frac{J}{2l}\\
J_{ADM} & = & \frac{M+1}{2}l+\frac{J}{2}\end{eqnarray*}
 which clearly violates the extremality condition $M_{ADM}^{2}l^{2}-J_{ADM}^{2}=-(Ml+J)l$
i.e. the regions having CTCs are naked since there are no horizons
to mask them. Clearly for such a spacetime causality breaks down.

\subsection{Left BTZ - Right zero geometries}

In this case we set, \[
A^{-}=0\]
 and keep the $A^{+}$ same as in the last case. So, the metric in
this case becomes \[
ds^{2}=\frac{Ml+J}{4}l(d\phi-\frac{dt}{l})^{2}+\frac{(1-\frac{Jl}{2r^{2}})^{2}}{4(N_{1}(r))^{2}}dr^{2}\]
 Clearly the metric determinant vanishes, it is non-invertible. So
these holomorphic geometries do not make sense either.

\section{Conclusion and Outlook}

If the path integral for pure gravity involves something like a holomorphically
factorized path integral over Chern-Simons gauge fields, we have seen
that large families of seemingly sensible gauge connections give rise
to geometries with naked closed timelike curves. Eliminating these
connections using some kind of physical state conditions that respect
holomorphicity would conversely eliminate the basic vacuum solution,
pure AdS. It seems then that physical state conditions must violate
holomorphicity, by for example, requiring that the inequality \eqref{eq:extremality}
be satisfied. Alternatively it may simply be the case that the Chern-Simons
formulation of pure gravity is not a good starting point for the quantization
of the theory. The study of limits of string theory compactifications
that give rise to pure gravity should provide useful clues to aid
further progress.

\begin{acknowledgments}
We thank Aaron Simons for helpful discussions. This research is supported
in part by DOE grant DE-FG02-91ER40688-Task A.
\end{acknowledgments}
\bibliographystyle{brownphys}
\bibliography{chiral-btz}

\end{document}